\documentclass[twocolumn]{aastex631}
\usepackage{graphicx}
\usepackage{amsmath}
\shorttitle{Impact of Upstream Clumpiness on SNR Evolution}
\shortauthors{TATSUMI \& INOUE}

\begin{document}
\title{Impact of Upstream Clumpiness on Supernova Remnant Forward Shock Evolution in Molecular Cloud Environments}
\author{Kenta Tatsumi}
\affiliation{Department of Physics, Konan University, Okamoto 8-9-1, Higashinada-ku, Kobe 658-8501, Japan}
\author{Tsuyoshi Inoue}
\affiliation{Department of Physics, Konan University, Okamoto 8-9-1, Higashinada-ku, Kobe 658-8501, Japan}
\email{tsuyoshi.inoue@konan-u.ac.jp}

\begin{abstract}
Supernova remnants (SNRs) are widely considered to be the primary accelerators of Galactic cosmic rays. In recent years, detailed observations have significantly progressed for young SNRs interacting with molecular clouds, a prime example being RX J1713.7$-$3946. When molecular clouds are clumpy, their impact can affect not only radiation properties but also shock wave propagation. Therefore, a quantitative understanding linking observational quantities with the ambient medium structure is highly required. In this study, we perform three-dimensional hydrodynamic simulations to model a molecular cloud with an inhomogeneous density structure driven by supersonic turbulence and subsequent SNR formation. To investigate various pre-supernova environments, we systematically vary the medium clumpiness by replacing gas below a threshold number density with a low-density hot gas, quantifying the relationship between the forward shock velocity and the volume filling factor of the high-density clumps. As a result, we find that at an elapsed time of $1000\ \mathrm{yr}$—a typical age for a young SNR—the forward shock can evolve consistently with the fast shock velocity measured in RX J1713.7$-$3946, provided that the clump volume filling factor is approximately $10\%$ or less. Considering that hadronic gamma-ray emission originates exclusively from the clumpy, high-density gas, our findings suggest that the total energy of cosmic-ray protons in RX J1713.7$-$3946 is higher than previously estimated, amounting to at least a few percent of the typical supernova explosion energy.
\end{abstract}


\section{Introduction}
Cosmic rays are high-energy non-thermal particles consisting primarily of nucleons, with protons being the dominant component.
In particular, Galactic cosmic rays with energies up to $3\ \mathrm{PeV}$ (the so-called knee energy) are widely believed to be accelerated at supernova remnant (SNR) shocks via the diffusive shock acceleration (DSA) mechanism \citep{Bell1978MNRAS182147,BlandfordOstriker1978ApJL221L29,BlandfordEichler1987PhR1541}.
Cosmic rays accelerated by SNRs emit electromagnetic radiations across a broad range of wavelengths, from radio to gamma-rays. Since the primary component of Galactic cosmic rays is protons, distinguishing between gamma-ray emission originating from accelerated electrons and that from protons is crucial for verifying cosmic-ray acceleration in SNRs. Electron-originated gamma-rays (leptonic gamma-rays) are produced via inverse Compton scattering, where cosmic-ray electrons scatter the cosmic microwave background photons, while proton-originated gamma-rays (hadronic gamma-rays) are generated when cosmic-ray protons collide with protons in the interstellar medium (ISM), producing neutral pions ($\pi^0$) that subsequently decay into gamma-ray photons. 

Hadronic gamma-ray emission is significantly enhanced by the interaction between an SNR and molecular clouds \citep{1996A&A...309..917A}. RX J1713.7$-$3946 \citep{1996rftu.proc..267P} is a prime example of such an SNR. It is the brightest SNR observed in the gamma-ray band, and its distance is estimated to be approximately $1\ \mathrm{kpc}$ based on the spatial correlation with molecular gas identified through CO observations \citep{Fukui2003PASJ55L61,Moriguchi2005ApJ631947}. It has been well established that if the ISM is uniformly distributed, the resulting hadronic and leptonic gamma-ray spectra differ substantially \citep{2006A&A...449..223A,Abdo2011ApJ73428}. While observations by the Fermi Gamma-ray Space Telescope reported a leptonic-like spectrum, subsequent studies have shown that if the molecular clouds have a clumpy distribution, the energy-dependent penetration of cosmic rays into high-density clumps makes the two emission mechanisms spectrally indistinguishable \citep{ZirakashviliAharonian2010ApJ708965,InoueYamazakiInutsukaFukui2012ApJ74471,GabiciAharonian2014MNRAS445L70,Inoue2019ApJ87246}.

In RX J1713.7$-$3946, magnetic field amplification up to $1\ \mathrm{mG}$ has been suggested by short-term variations in X-ray emission \citep{Uchiyama2007Nature449576}. Such strong magnetic field amplification can be explained by a turbulent dynamo operating during the interaction between a clumpy molecular cloud and a supernova shock wave \citep{InoueYamazakiInutsuka2009ApJ695825,InoueYamazakiInutsukaFukui2012ApJ74471,Sano2013ApJ77859}. Furthermore, recent observational analyses correlating gamma-ray and X-ray data with the interstellar proton distribution have estimated that approximately 70\% of the gamma-ray emission from RX J1713.7$-$3946 is hadronic, with the remaining 30\% being leptonic \citep{Fukui2021ApJ91584}. Since several other young SNRs exhibit similar characteristics \citep{2014ApJ...788...94F,2019ApJ...876...37S,2024ApJ...961..162F}, understanding the impact of shock-cloud interactions on SNR evolution is of paramount importance.

A clumpy medium can influence not only the radiation properties but also the propagation of the shock wave itself. However, it remains poorly quantified how the density distribution and the volume filling factor of high-density clumps affect shock propagation, and how these effects propagate into observational estimates of cosmic-ray acceleration.

In this study, we perform three-dimensional hydrodynamic simulations of SNR formation, adopting an inhomogeneous molecular cloud shaped by supersonic turbulence as the initial ISM state. To investigate how the evolution depends on the upstream environment, we systematically vary the volume filling factor of the high-density clumps by replacing gas below a certain threshold number density with low-density hot gas.
Such an environment, where high-density clumps are embedded in a hot gas while middle-density components have already been blown away, is expected to form naturally around massive stars due to stellar winds and the expansion of H II regions \citep{InoueYamazakiInutsukaFukui2012ApJ74471}. In this picture, whether a clump survives or is cleared out depends on its density, which determines our model setup described in Section 2.2.

This paper is organized as follows. In Section~2, we describe the numerical setup for our simulations, and in Section~3, we present the simulation results. Based on these results, we discuss the necessary corrections for the observational estimation of cosmic-ray energy in SNR RX J1713.7$-$3946 in Section~4. Finally, we summarize our findings in Section~5.

\section{Method and Procedure}

We perform three-dimensional hydrodynamic simulations of inhomogeneous interstellar gas generation and subsequent SNR formation using the astrophysical fluid dynamics code Athena++ \citep{Stone2020ApJS2494}. Athena++ employs the HLLC approximate Riemann solver \citep{1994ShWav...4...25T} to integrate the compressible hydrodynamic equations, combined with a second-order spatial piecewise linear method and a second-order Runge-Kutta time integration scheme. The governing equations are expressed as follows:
\begin{equation}
  \frac{\partial \rho} {\partial t} + \mathbf{\nabla} \cdot (\rho \mathbf{v}) = 0,
  \label{eq:continuity}
\end{equation}
\begin{equation}
  \frac{\partial (\rho \mathbf{v})} {\partial t} + \mathbf{\nabla} (\rho \mathbf{v} \mathbf{v}^{T} + p I) = 0,
  \label{eq:momentum_conservation}
\end{equation}
\begin{equation}
  \frac{\partial E} {\partial t} + \mathbf{\nabla} \cdot [(E + p) \mathbf{v}] = 0,
  \label{eq:energy_conservation}
\end{equation}
where $I$ denotes the identity matrix, and $\rho$, $p$, and $\mathbf{v}$ represent the density, pressure, and velocity, respectively. The total energy density is given by $E = e + \rho v^2 / 2$, where $e = p / (\gamma - 1)$ is the internal energy density.

In this paper, the simulations are divided into two parts. First, we perform a simulation to generate a molecular cloud with an inhomogeneous density structure driven by supersonic turbulence. Subsequently, we initialize the SNR ejecta inside the generated molecular cloud to simulate the formation and evolution of the SNR.

\subsection{Generation of Inhomogeneous Structures in Molecular Clouds}

Molecular clouds are observationally known to always feature supersonic turbulence. This supersonic turbulence drives shocks that create clumpy and anisotropic structures inside the molecular clouds\footnote{Simulations of molecular cloud formation have shown that molecular clouds already exhibit clumpy distributions even from the stage of their parental \ion{H}{1} clouds due to the thermal instability \citep{InoueInutsuka2008ApJ687303,InoueInutsuka2012ApJ75935}. Therefore, a molecular cloud cannot be considered uniform at any stage of its evolution.}. In this section, we present the formation process of such structures through numerical simulations. The numerical domain is a cubic box with a side length of $L_{\rm box}= 10\ \mathrm{pc}$ and a resolution of $512^3$ grid cells. Because molecular clouds are isothermal systems where radiative cooling is highly efficient, we approximate the soft equation of state of the molecular cloud at this stage by setting the specific heat ratio to $\gamma = 1.05$ \citep[e.g.,][]{2000ApJ...538..115S}.
The mean molecular weight of the gas is set to $m = 2.4\,m_{\rm p}$.

A spherical molecular cloud with a radius of $3\ \mathrm{pc}$, a number density of $n_{\rm MC} = 10^3\ \mathrm{cm^{-3}}$, and a temperature of $T_{\rm MC} = 10\ \mathrm{K}$ is placed at the center of the computational domain. The cloud is surrounded by a diffuse warm gas with a number density of $n_{\rm W} = 1\ \mathrm{cm^{-3}}$ and a temperature of $T_{\rm W} = 10^4\ \mathrm{K}$. In this initial configuration, the molecular cloud sphere and the diffuse ISM are in pressure equilibrium. The total mass enclosed within the numerical domain is $6.8\times10^3\ M_\odot$. Periodic boundary conditions are applied to all boundaries.

To generate anisotropic structures, a turbulent velocity field is applied across the entire computational domain. The turbulence is driven by a stochastic external force generated in Fourier space, with the driving wavenumbers set to $k = (1-3) \times 2\pi/L_{\rm box}$. The external forcing is updated with a correlation time of $t_{\rm corr} = 1\ \mathrm{Myr}$, and the amplitude of the turbulent field is scaled so that the turbulent velocity dispersion corresponds to a Mach number of $M = 10$ for 10 K molecular gas. Under these conditions, the turbulent crossing time is $5.28\ \mathrm{Myr}$. We continue the simulation up to $10.55\ \mathrm{Myr}$, which is approximately twice the crossing time.

\begin{figure}[htbp]
  \centering
  \includegraphics[keepaspectratio, scale=0.3]{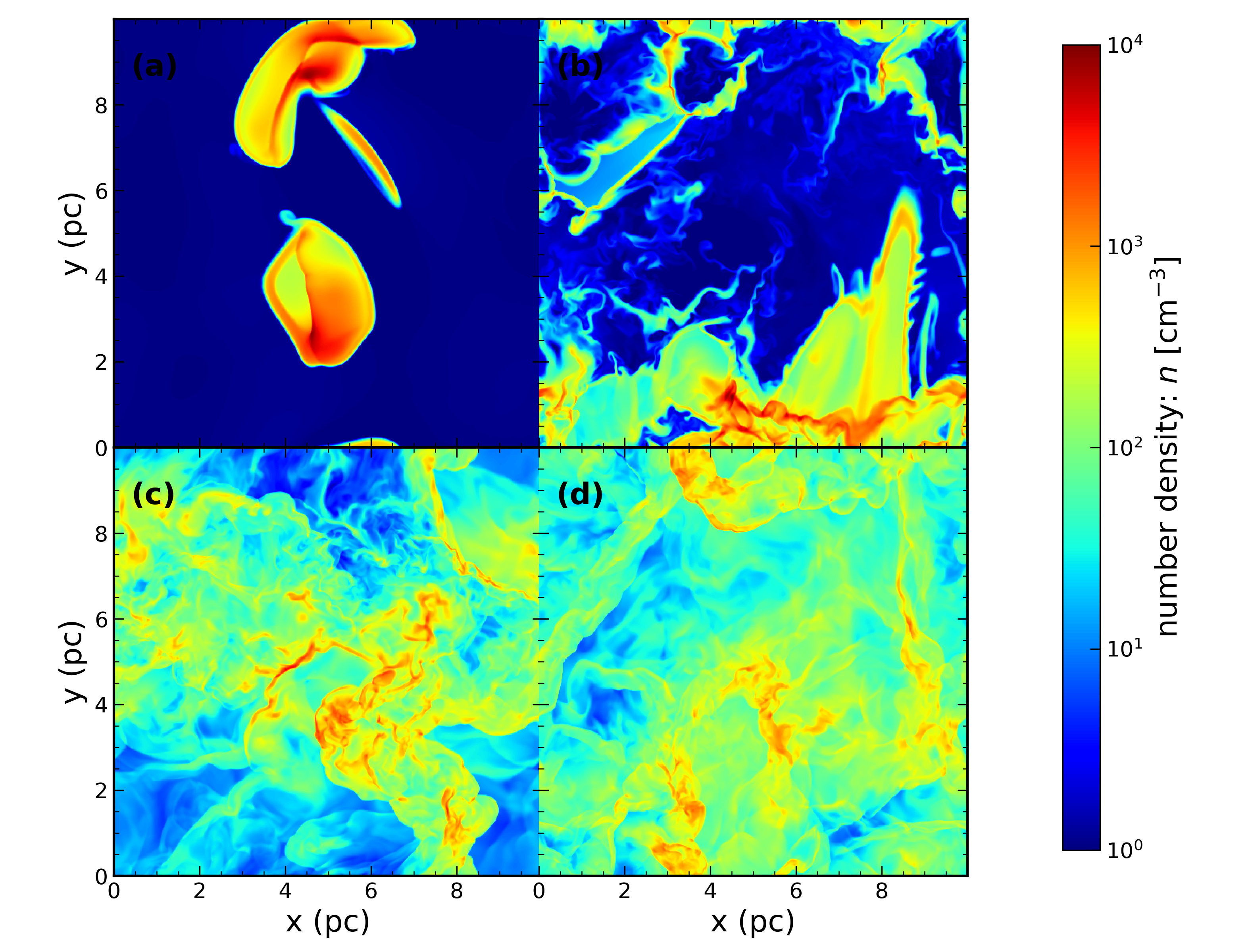}
  \caption{Slice plots of the number density in the turbulent molecular gas at the $z = 5.0\ \mathrm{pc}$ plane. The panels show the states at $t = 2.64\ \mathrm{Myr}$ (top-left), $5.28\ \mathrm{Myr}$ (top-right), $7.92\ \mathrm{Myr}$ (bottom-left), and $10.55\ \mathrm{Myr}$ (bottom-right).}
  \label{fig:mcloud_number_density_slices_timeseries}
\end{figure}

\begin{figure}[htbp]
  \centering
  \includegraphics[keepaspectratio, scale=0.5]{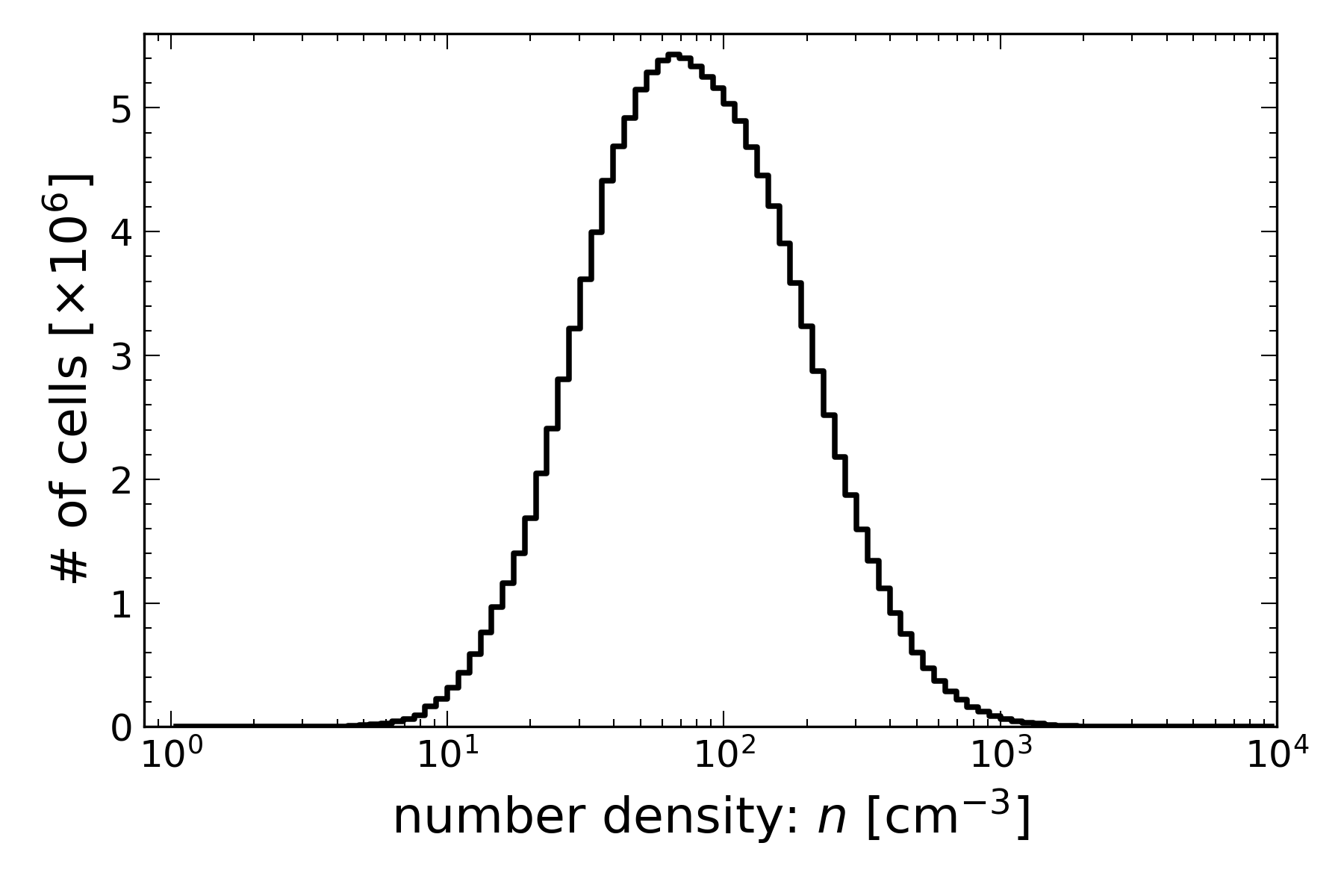}
  \caption{Number density probability distribution histogram at $t = 10.55\ \mathrm{Myr}$.}
  \label{fig:number_density_histogram_5myr}
\end{figure}

Figure~\ref{fig:mcloud_number_density_slices_timeseries} shows the time evolution of the number density slice map at the $z = 5.0\ \mathrm{pc}$ plane. Panels (a)--(d) present the results at $t = 2.64$, $5.28$, $7.92$, and $10.55\ \mathrm{Myr}$, respectively. It is clearly seen that by two eddy turnover times ($t = 10.55\ \mathrm{Myr}$), the structure has sufficiently evolved due to the supersonic turbulence, resulting in the formation of numerous high-density clumps. Figure~\ref{fig:number_density_histogram_5myr} shows the number density probability distribution histogram at $t = 10.55\ \mathrm{Myr}$. The number density distribution follows a well-known lognormal distribution \citep{PassotVazquezSemadeni1998,2009A&A...508L..35K}, with a mean number density of $114\ \mathrm{cm^{-3}}$ and a standard deviation of 0.47~dex.

\subsection{Simulation of SNR Formation}

As the initial conditions for this stage, we utilize the inhomogeneous density structure obtained in the previous section. While the mean molecular weight of the gas is set to $2.4\,m_{\rm p}$ in the previous section, the shock-heated molecular gas is expected to dissociate and subsequently ionize into a plasma. Therefore, in this section and hereafter, we adopt the nucleon number density defined as $n = \rho / m_{\rm p}$, rather than the particle number density. For simplicity, we assume an initial pressure of $p/k_{\rm B}=10^4\ \mathrm{[K\ cm^{-3}]}$, which is a typical value for molecular clouds. Since this pressure is sufficiently small compared to the pressure inside the supernova remnant, the simulation results do not depend sensitively on the exact choice of this value.

In the ISM surrounding a massive star, the low-density component is expected to be cleared out by effects such as the expansion of \ion{H}{2} regions and stellar winds.
The extent to which these low-density regions are swept up highly depends on the properties of the nearby massive star. Therefore, in this paper, we model the pre-supernova environment by replacing the gas below a certain threshold number density ($n_{\rm th}$) with a low-density gas of $n_{\rm hot} = 0.05\ \mathrm{cm^{-3}}$. This specific density ($n_{\rm hot}$) satisfies the observational constraints on thermal X-ray emission from RX J1713.7$-$3946 \citep{2008PASJ...60S.131T}. We examine seven different threshold value cases of $n_{\rm th} = 0, 100$, $250$, $500$, $750$, $1000\ \mathrm{cm^{-3}}$, and $\infty$, where $n_{\rm th} = 0$ and $\infty$ means no and full replacement to the hot gas, respectively.

\begin{figure}[htbp]
  \centering
  \includegraphics[keepaspectratio, scale=0.6]{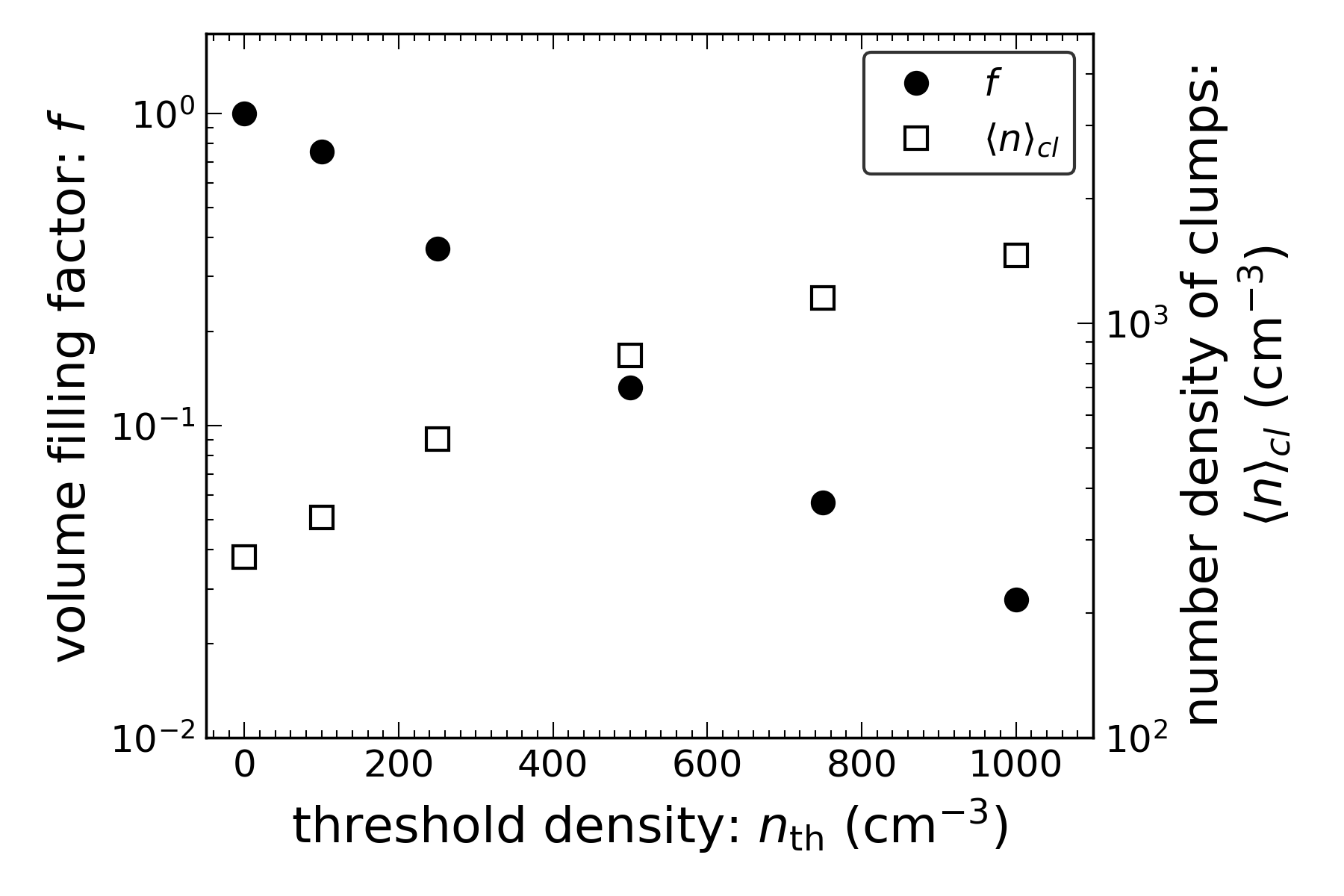}
  \caption{Relationships among the threshold number density, the mean number density of the high-density clumps, and the volume filling factor.}
  \label{fig:nth_vs_nclmean_vff}
\end{figure}

The replacement of the low-density gas mentioned above renders the ambient medium more clumpy.
The resulting medium can be well characterized by the volume filling factor ($f$) of the dense clumps.
The relation between the threshold density $n_{\rm th}$ and the volume filling factor $f$ is shown in Figure~\ref{fig:nth_vs_nclmean_vff}.
Since a higher threshold density leads to a larger mean density of the remaining clumps ($\langle n \rangle_{\rm cl}$), we also plot $\langle n \rangle_{\rm cl}$ as a function of $n_{\rm th}$ for reference in the same figure.

To simulate the formation of the SNR, a spherical ejecta component is initialized within the numerical domain. To maximize computational efficiency, we model the first octant rather than the full sphere, placing the center of the octant at one corner of the domain ($x = y = z = 0\ \mathrm{pc}$).
Note that the numerical domain (including the origin) is identical to that used in the previous section.
The mass of the injected ejecta within this octant is $0.504\ M_\odot$ (corresponding to $4.03\ M_\odot$ for the full sphere). The initial radius of the ejecta is set to $0.5\ \mathrm{pc}$, and the velocity profile within the ejecta is assumed to be linear:
\begin{equation}
  \mathbf{v} = v_{\rm max} \frac{\mathbf{r}}{r_{\rm max}},
  \label{eq:linear_velocity_profile}
\end{equation}
following \citet{Chevalier1982ApJ258790} and \citet{Petruk2021MNRAS505755}. Here, $r_{\rm max} = 0.5\ \mathrm{pc}$ and $v_{\rm max} = 10^4\ \mathrm{km\ s^{-1}}$, which corresponds to a total kinetic energy of $2.4 \times 10^{51}\ \mathrm{erg}$ for the full sphere.

Reflecting boundary conditions are applied to the boundaries adjacent to the ejecta, while free-outflow boundary conditions are used for the remaining outer boundaries. The simulation is integrated for a duration of $t_{\rm dur} = 10^3\ \mathrm{yr}$, which is comparable to the crossing time of the resulting supernova shock wave. Furthermore, because this timescale is shorter than the typical cooling timescale of the bulk plasma, we perform the simulations under adiabatic conditions with a specific heat ratio of $\gamma = 5/3$. Note that although the cooling timescale in the shocked dense gas can be as short as $t_{\rm dur}$, radiative cooling in these dense regions would not significantly alter the global shock propagation velocity, which is the primary focus of the discussion below.

\section{Result}

\subsection{Overview}

\begin{figure*}[htbp]
  \centering
  \includegraphics[
    scale=0.65,
    keepaspectratio
  ]{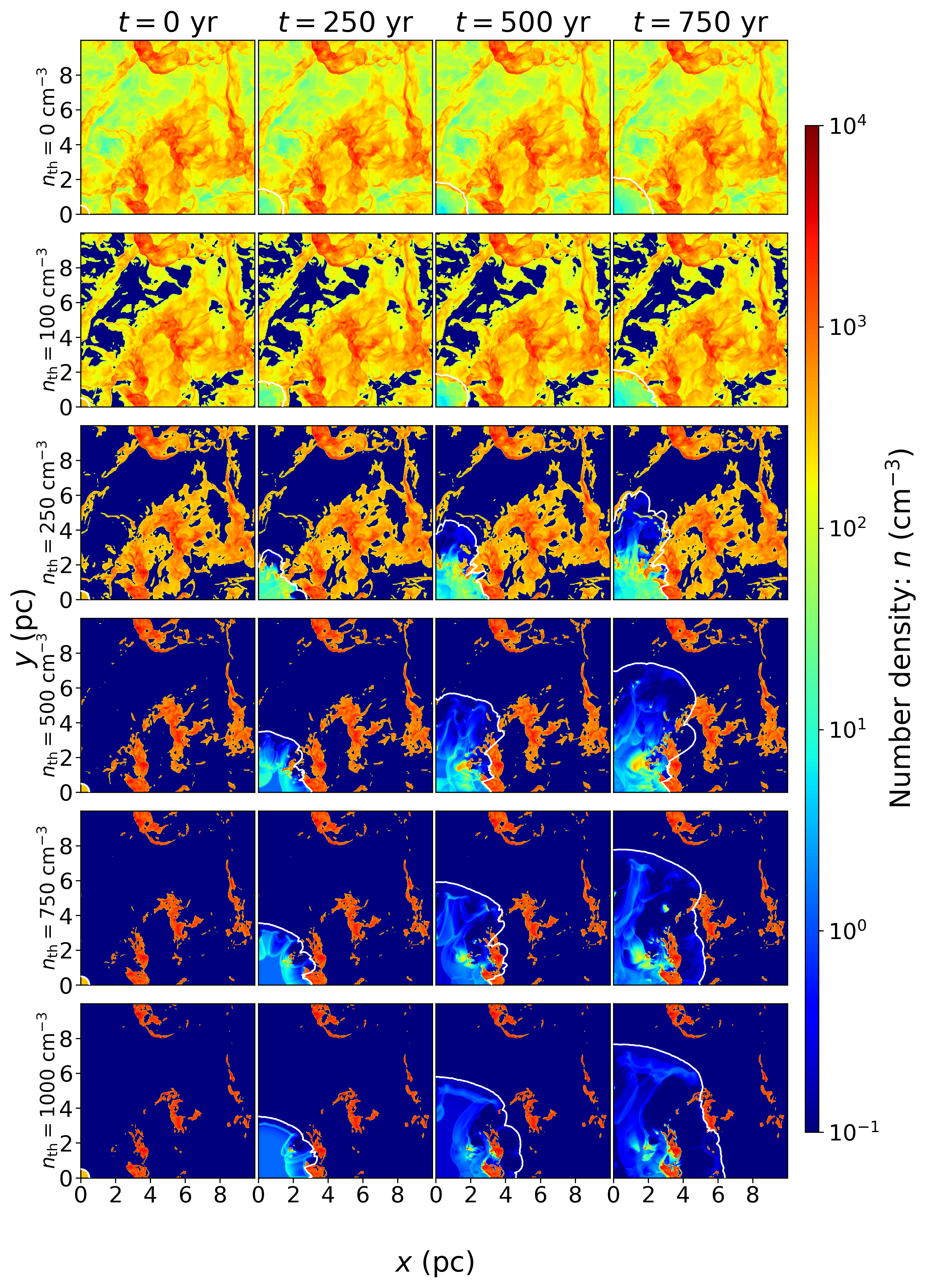}
  \caption{Time evolution of the gas number density at the $z=0.0$ pc plane in the simulations with various threshold number densities ($n_{\rm th}$). The columns from left to right correspond to $t = 0$, $250$, $500$, and $750\ \mathrm{yr}$, while the rows from top to bottom correspond to $n_{\rm th} = 0$, $100$, $250$, $500$, $750$, and $1000\ \mathrm{cm^{-3}}$.
  The white lines indicate the forward shock fronts.
  }
  \label{fig:snr_n_timeseries_nth}
\end{figure*}

\begin{figure}[htbp]
  \centering
  \includegraphics[
    width=\linewidth,
    keepaspectratio
  ]{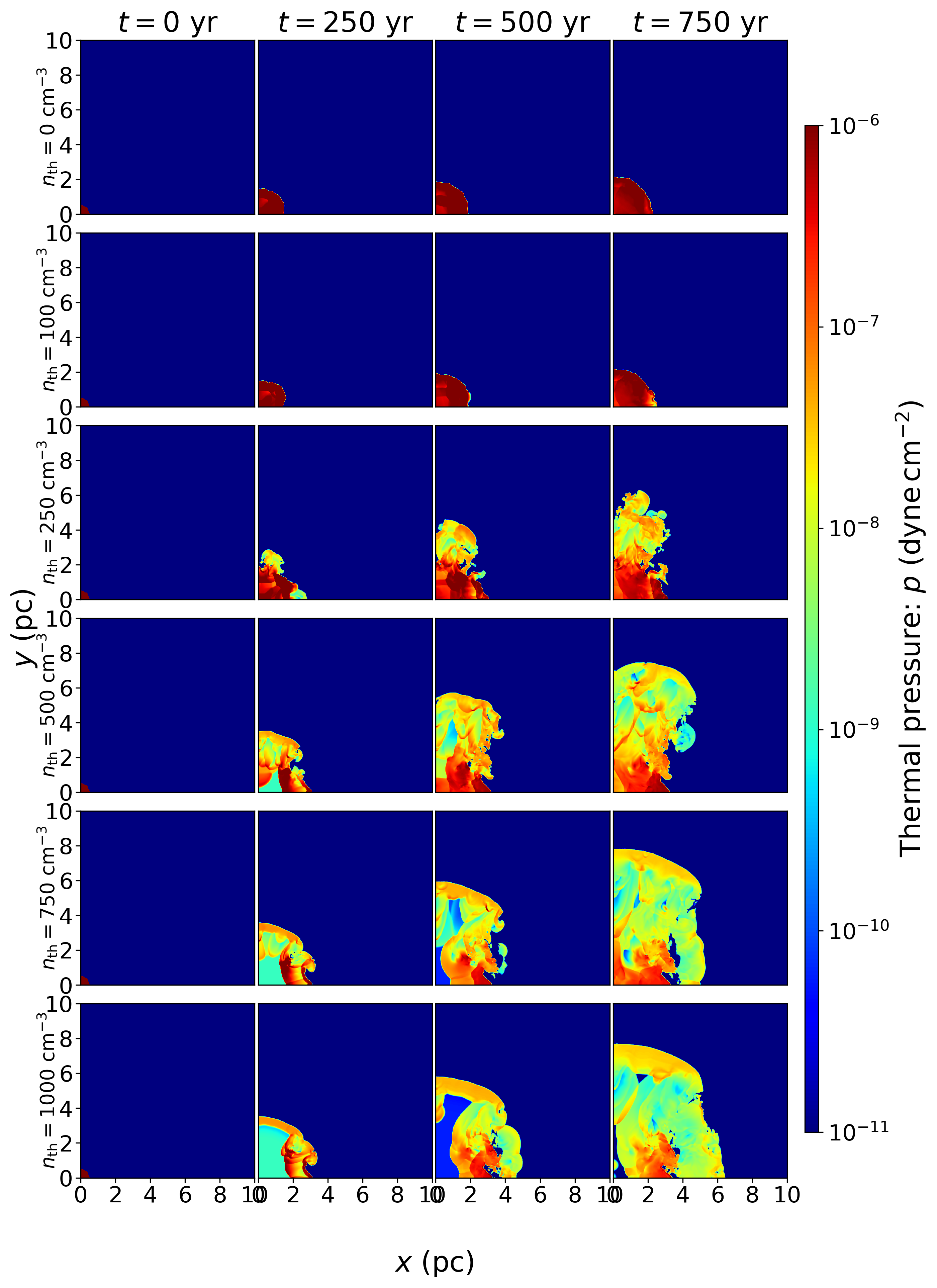}
  \caption{Same as Figure~\ref{fig:snr_n_timeseries_nth}, but for the thermal pressure $p$.}
  \label{fig:snr_p_timeseries_nth}
\end{figure}

Figure~\ref{fig:snr_n_timeseries_nth} compares the time evolution of the number density distribution for the six runs with different threshold number densities ($n_{\rm th}$). In this figure, the rows from top to bottom correspond to $t = 0$, $250$, $500$, and $750\ \mathrm{yr}$, while the columns from left to right correspond to $n_{\rm th} = 0$, $100$, $250$, $500$, $750$, and $1000\ \mathrm{cm^{-3}}$ cases.
White lines indicate forward shock position (the method for identifying the shock front is described in the next section).
Figure~\ref{fig:snr_p_timeseries_nth} presents the same panel layout as Figure~\ref{fig:snr_n_timeseries_nth}, but displays the thermal pressure $p$.
In the case of $n_{\rm th} = 0\ \mathrm{cm^{-3}}$, because the ambient gas density is high throughout the domain, the shock wave immediately enters the deceleration phase. Consequently, the radius of the SNR reaches an average of only $2.3\ \mathrm{pc}$ even after $1000\ \mathrm{yr}$.

For cases with $n_{\rm th} > 500\ \mathrm{cm^{-3}}$, the SNR propagates at a significantly higher velocity than in the $n_{\rm th} = 0\ \mathrm{cm^{-3}}$ case by passing through the clearing between the high-density clumps. As predicted by \citet{InoueYamazakiInutsukaFukui2012ApJ74471}, our results support that the highly spherical morphology of the emission from RX J1713.7$-$3946 can be explained by the SNR shock wave expanding through the inter-clump spaces.

For the models with $n_{\rm th} \ge 250$ cm$^{-3}$, the thermal pressure distributions exhibit highly filled structures of high-pressure regions, which contrast with the localized and inhomogeneous structures seen in the gas number density distributions. This behavior can be understood as a result of multiple shock reflections. When the forward shock collides with dense, rigid clumps, it generates strong reflected and bow shocks \citep{2010ApJ...723L.108I}. These secondary shocks repeatedly interact and propagate through the low-density inter-clump medium, efficiently converting the kinetic energy of the bulk flow into thermal energy. Consequently, while the bulk of the mass remains confined within the dense clumps, the inter-clump space becomes uniformly pressurized and filled with high-pressure gas.

\subsection{Dependence of the Shock Velocity on the Volume Filling Factor.}

\begin{figure}[htbp]
  \centering
  \includegraphics[keepaspectratio, scale=0.5]{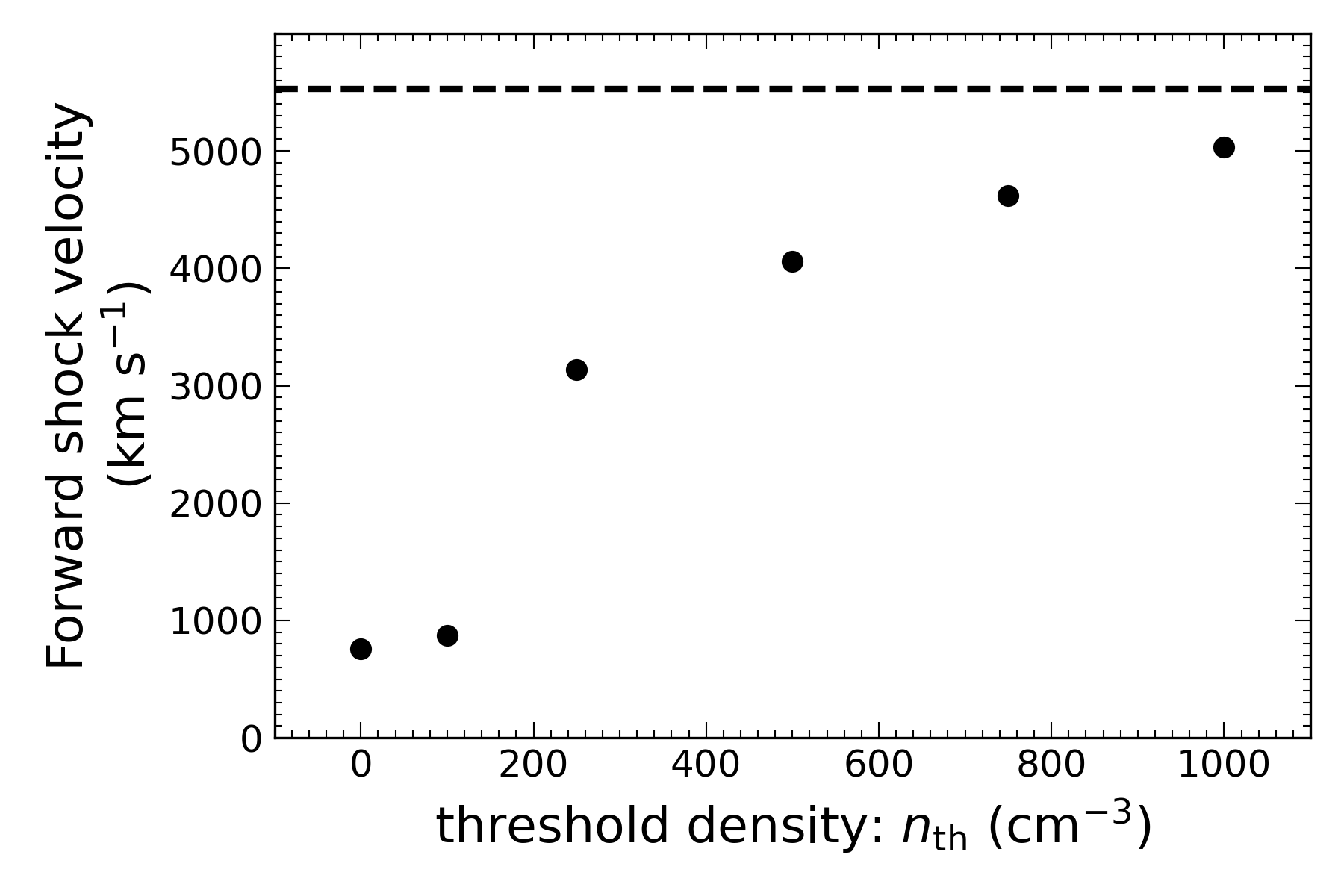}
  \caption{Mean forward shock velocity vs. threshold number density at $t = 1000\ \mathrm{yr}$. The dashed line represents the shock velocity for the case with $n_{\rm th} = \infty$.}
  \label{fig:nth_vs_mean_forward_shock_velocity}
\end{figure}

Figure~\ref{fig:nth_vs_mean_forward_shock_velocity} shows the relationship between the threshold number density and the mean forward shock velocity at $t = 1000\ \mathrm{yr}$. Here, the forward shock velocity is measured as follows: we define a spherical coordinate system $(r, \theta, \phi)$ with a common origin to the numerical domain. Then, the thermal pressure $p$ is scanned in the negative radial direction from the outer boundary of the SNR, and the position where $p \ge 10^5\,k_{\mathrm{B}}\ \mathrm{[K\ cm^{-3}]}$ is identified as the shock front. This procedure is applied along $100 \times 100$ directions covering $(\theta, \phi)$ to map the entire shock front. The mean radial velocity of the shock wave is then calculated from the displacement of the shock front position sampled every $10\ \mathrm{yr}$. The dashed line in Figure~\ref{fig:nth_vs_mean_forward_shock_velocity} represents the forward shock velocity for the case with $n_{\rm th} = \infty$ (i.e., where the entire region outside the ejecta is filled with the hot gas of $n_{\rm hot} = 0.05\ \mathrm{cm^{-3}}$). The mean shock velocity exhibits a distinct transition around $n_{\rm th} \sim 200\ \mathrm{cm^{-3}}$; for cases with $n_{\rm th} \le 200\ \mathrm{cm^{-3}}$, the forward shock velocity drops significantly below $3000\ \mathrm{km\ s^{-1}}$.

\begin{figure}[htbp]
  \centering
  \includegraphics[keepaspectratio, scale=0.5]{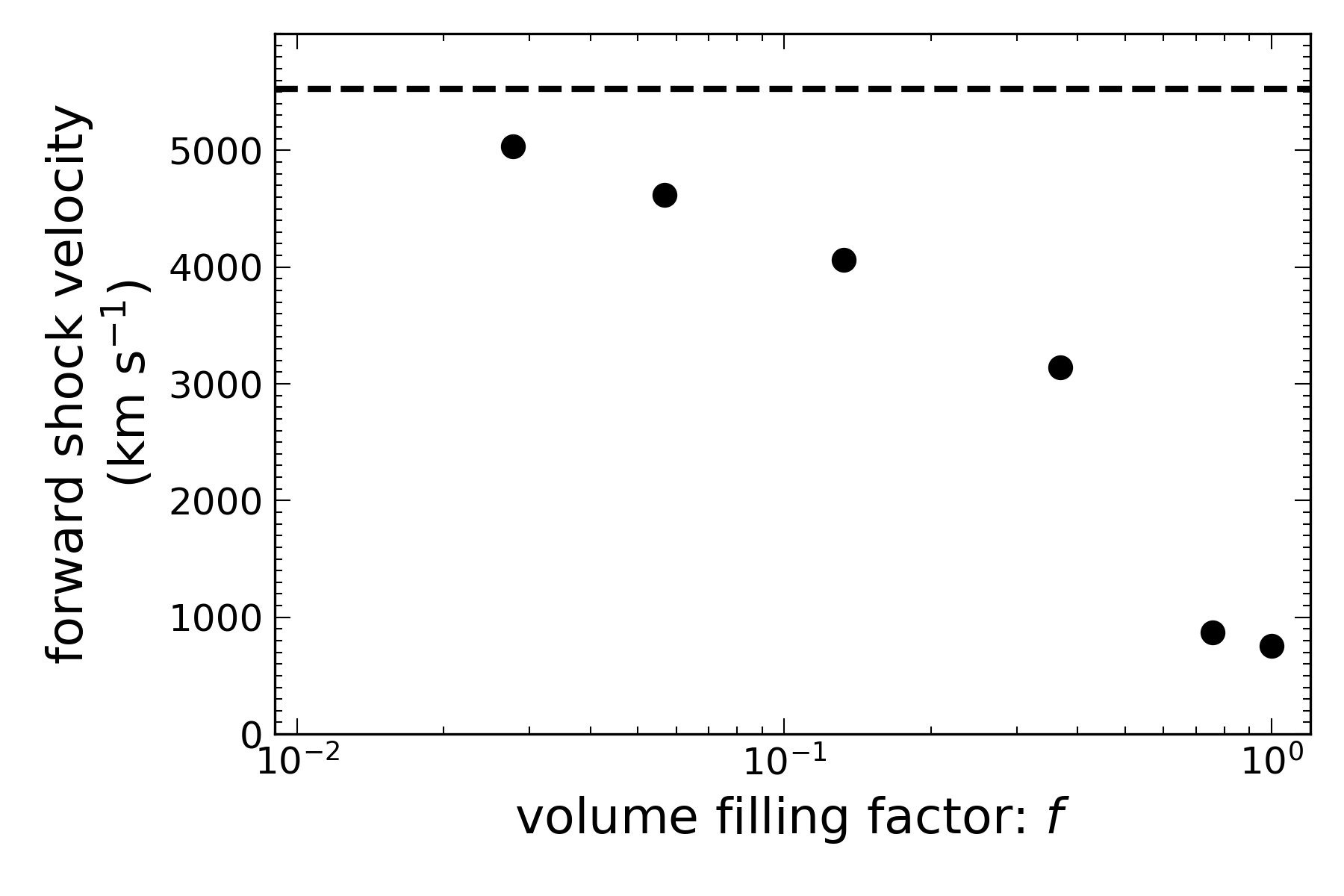}
  \caption{Mean forward shock velocity vs. volume filling factor at $t = 1000\ \mathrm{yr}$. The dashed line represents the shock velocity for the case with $n_{\rm th} = \infty$.}
  \label{fig:vff_vs_mean_forward_shock_velocity}
\end{figure}

Figure~\ref{fig:vff_vs_mean_forward_shock_velocity} is identical to Figure~\ref{fig:nth_vs_mean_forward_shock_velocity}, except that the horizontal axis represents the initial volume filling factor $f$ of the high-density clumps with densities greater than $n_{\rm th}$. Here, the volume filling factor is defined as the ratio of the number of grid cells exceeding the threshold density to the total number of cells in the computational domain. It is clearly seen that when the volume filling factor significantly exceeds 10\%, the forward shock velocity decreases substantially compared to the $n_{\rm th} = \infty$ case.

\begin{figure}[htbp]
  \centering
  \includegraphics[keepaspectratio, scale=0.5]{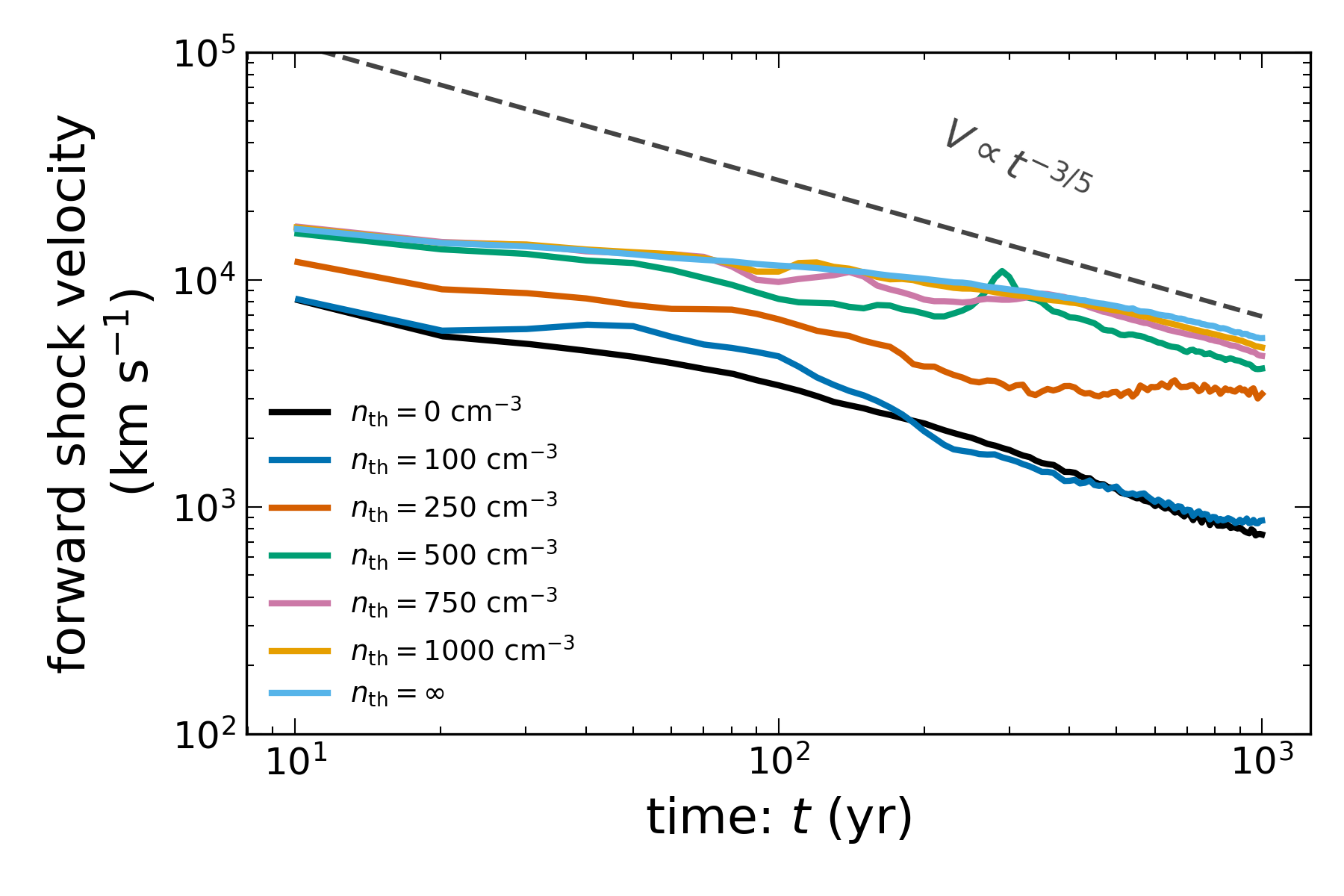}
  \caption{Time evolution of the forward shock velocity ($v_{\rm sh}$) for models with various threshold densities ($n_{\rm th}$). The black, blue, red, green, pink, orange, and light blue lines correspond to models with $n_{\rm th} = 0$, $100$, $250$, $500$, $750$, $1000$, and $\infty\ \mathrm{cm^{-3}}$, respectively. The dashed line indicates the time dependence of the shock velocity expected from the Sedov solution ($v_{\rm sh} \propto t^{-3/5}$).}
  \label{fig:shock_speed_vs_time}
\end{figure}

\begin{figure}[htbp]
  \centering
  \includegraphics[keepaspectratio, scale=0.5]{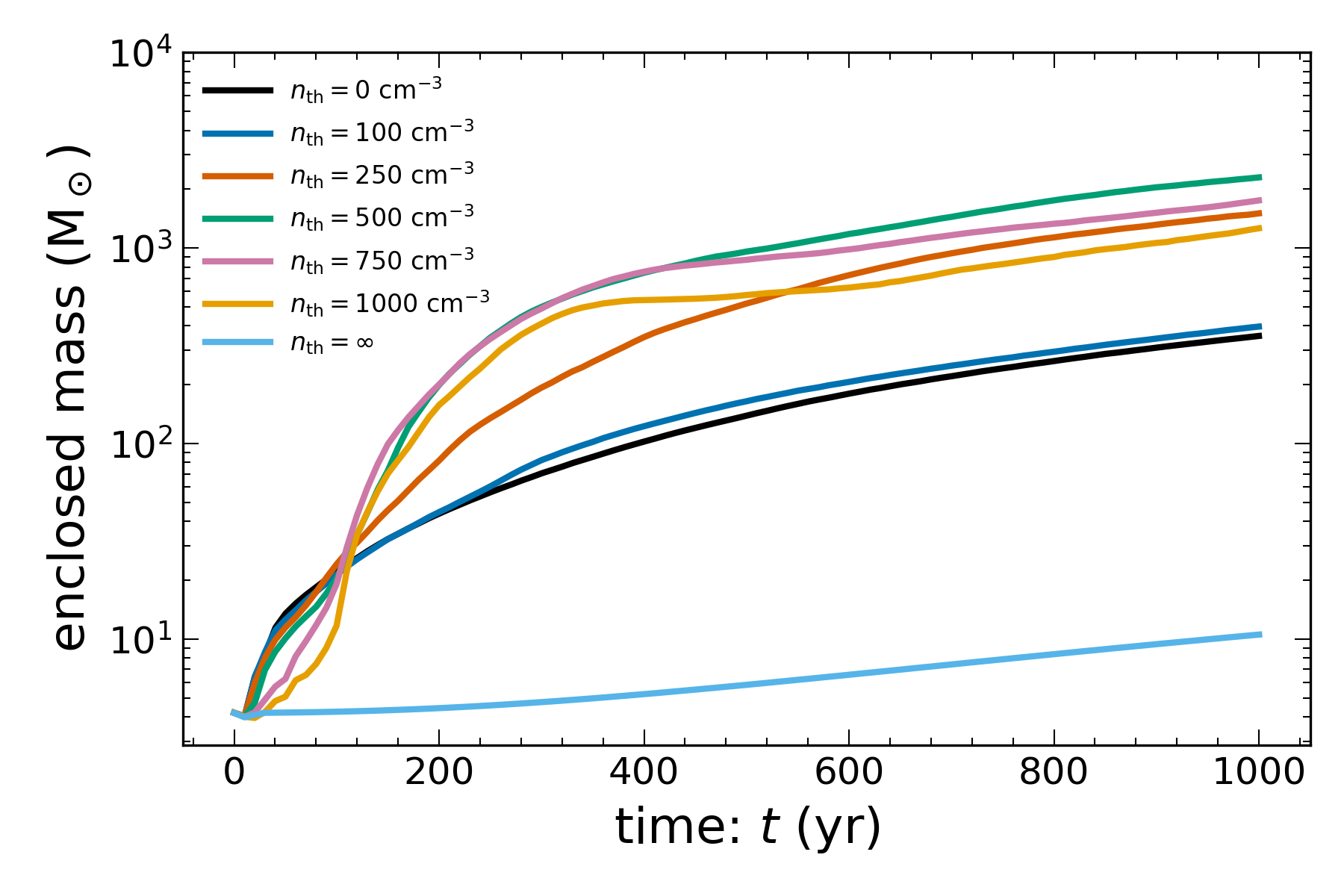}
  \caption{
Time evolution of the gas mass inside the shock wave for models with various threshold densities ($n_{\rm th}$).
  The black, blue, red, green, pink, orange, and light blue lines correspond to models with $n_{\rm th} = 0$, $100$, $250$, $500$, $750$, $1000$, and $\infty\ \mathrm{cm^{-3}}$, respectively.}
  \label{fig:snr_shell_mass_vs_time}
\end{figure}

Figure~\ref{fig:shock_speed_vs_time} illustrates the time evolution of the shock velocity. In the case of $n_{\rm th} = \infty$, the deceleration phase (Sedov phase) begins at around $t \sim 1000\ \mathrm{yr}$. This onset is naturally expected because the mass of the hot gas swept up within the shock radius ($r_{\rm sh} \sim 8$--$9\ \mathrm{pc}$) at this epoch becomes comparable to the mass of the initialized ejecta. Figure~\ref{fig:snr_shell_mass_vs_time} shows the evolution of the gas mass inside the shock wave for each model.
Interestingly, in the cases with $f < 0.1$ ($n_{\rm th} > 750\ \mathrm{cm^{-3}}$), the evolutionary profiles of the shock velocity do not differ significantly from that of the $n_{\rm th} = \infty$ case. However, the enclosed mass at $t = 1000\ \mathrm{yr}$ exceeds $10^3\ M_\odot$ even in the $n_{\rm th} = 1000\ \mathrm{cm^{-3}}$ case. It is well known that a blast wave enters its deceleration phase when the swept-up gas mass becomes comparable to the ejecta mass. Nevertheless, when the ISM is clumpy and its volume filling factor is less than approximately 10\%, the transition to the deceleration phase is significantly delayed compared to this conventional expectation. This indicates that for $f < 0.1$, the shock wave can propagate through the gaps between the dense clumps, meaning that only the diffuse inter-clump gas contributes to the immediate deceleration of the shock front.

In Figure~\ref{fig:shock_speed_vs_time}, the model with $n_{\rm th} = 500\ \mathrm{cm^{-3}}$ ($f \simeq 0.1$) exhibits a bump in the shock velocity at around $t = 300\ \mathrm{yr}$, which can be attributed to the shock front passing through the large clumps, i.e., a shock breakout into the inter-clump medium.
The fact that this model yields the maximum enclosed mass in Figure~\ref{fig:snr_shell_mass_vs_time} is a natural consequence, as it represents the medium with the highest possible volume filling factor of clumps that can exist without causing a severe deceleration of the shock wave.

As for the forward shock velocity of RX J1713.7$-$3946, a velocity of $(3900 \pm 300)(d / 1\ \mathrm{kpc})\ \mathrm{km\ s^{-1}}$ has been reported for the northwestern (NW) shell \citep{TsujiUchiyama2016PASJ68108}. Additionally, at the western edge of the southwestern (SW) rim, proper motion measurements have yielded a velocity of $(3800 \pm 100)(d / 1\ \mathrm{kpc})\ \mathrm{km\ s^{-1}}$ \citep{Tanaka2020ApJL900L5}. Although the initial explosion velocity cannot be directly constrained, it is physically unlikely that the shock wave has decelerated to less than 30\% of its initial velocity as a whole.

Figure~\ref{fig:vff_vs_mean_forward_shock_velocity} demonstrates that when the volume filling factor of the dense clumps exceeds roughly 10\%, the shock velocity is decelerated to less than 30\% of the initial explosion velocity. Consequently, our findings strongly suggest that the ambient environment surrounding RX J1713.7$-$3946 must have a high-density clump volume filling factor of 10\% or less.

\section{Discussion}

\subsection{Total Energy of Cosmic-Ray Protons}

Based on radio and gamma-ray observations of the ISM surrounding RX J1713.7$-$3946, \citet{Fukui2012ApJ74682} estimated the total energy of cosmic-ray protons in RX J1713.7$-$3946 as follows:
\begin{equation}
  W_{\rm tot} = (2 \pm 1) \times 10^{48} \left(\frac{d}{1\ \mathrm{kpc}}\right) \left(\frac{n}{100\ \mathrm{cm^{-3}}} \right)^{-1}\ \mathrm{erg},
  \label{eq:wtot_scaling_d_n}
\end{equation}
where $n$ represents the mean density of the surrounding ISM.

Our simulation results demonstrate that the ambient ISM surrounding RX J1713.7$-$3946 must be clumpy, with a volume filling factor of approximately 0.1 or less. If this is the case, the amount of target gas decreases, and the estimation in Equation~(\ref{eq:wtot_scaling_d_n}) should be modified as follows:
\begin{equation}
  W_{\rm tot} = (2 \pm 1) \times 10^{49} \left(\frac{d}{1\ \mathrm{kpc}}\right) \left(\frac{\langle n\rangle_{\rm cl}}{100\ \mathrm{cm^{-3}}} \right)^{-1}\left(\frac{f}{0.1} \right)^{-1}\ \mathrm{erg},
  \label{eq:Wtot2}
\end{equation} 
where $f$ is the volume filling factor of the clumps.

\begin{figure}[htbp]
  \centering
  \includegraphics[keepaspectratio, scale=0.5]{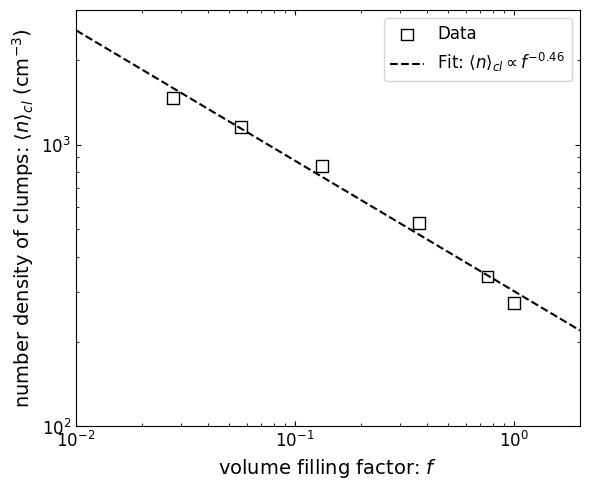}
  \caption{Relationship between the volume filling factor and the mean density of the clumps. The dashed line represents the best-fit result using the least-squares method, demonstrating that the mean clump density is well approximated by the relation $\langle n\rangle_{\rm cl} \propto f^{-0.46}$.}
  \label{fig:ncl_vs_f}
\end{figure}

In our initial setup, specifying $n_{\rm th}$ determines both the mean number density of the clumps, $\langle n\rangle_{\rm cl}$, and the volume filling factor, $f$.
Figure~\ref{fig:ncl_vs_f} illustrates this relationship. Fitting the data using the least-squares method shows that the mean density of the clumps is well approximated by the relation $\langle n\rangle_{\rm cl} \propto f^{-0.46}$. For RX J1713.7$-$3946, it is expected that $\langle n\rangle_{\rm cl} \simeq 100 \times f^{-0.46}\ \mathrm{cm^{-3}}$. Utilizing this relation to eliminate $\langle n\rangle_{\rm cl}$ from Equation~(\ref{eq:Wtot2}) yields
\begin{equation}
  W_{\rm tot} = (0.69 \pm 0.35) \times 10^{49} \left(\frac{d}{1\ \mathrm{kpc}}\right) \left(\frac{f}{0.1} \right)^{-0.54}\ \mathrm{erg}.
  \label{eq:Wtot3}
\end{equation} 
This implies that more than approximately 1\% of the typical supernova explosion energy ($10^{51}\ \mathrm{erg}$) is channeled into cosmic-ray acceleration, which is fully consistent with conventional expectations for acceleration efficiency.

\subsection{Mass Distribution around RX J1713.7$-$3946}
\begin{figure}[htbp]
  \centering
  \includegraphics[keepaspectratio, scale=0.5]{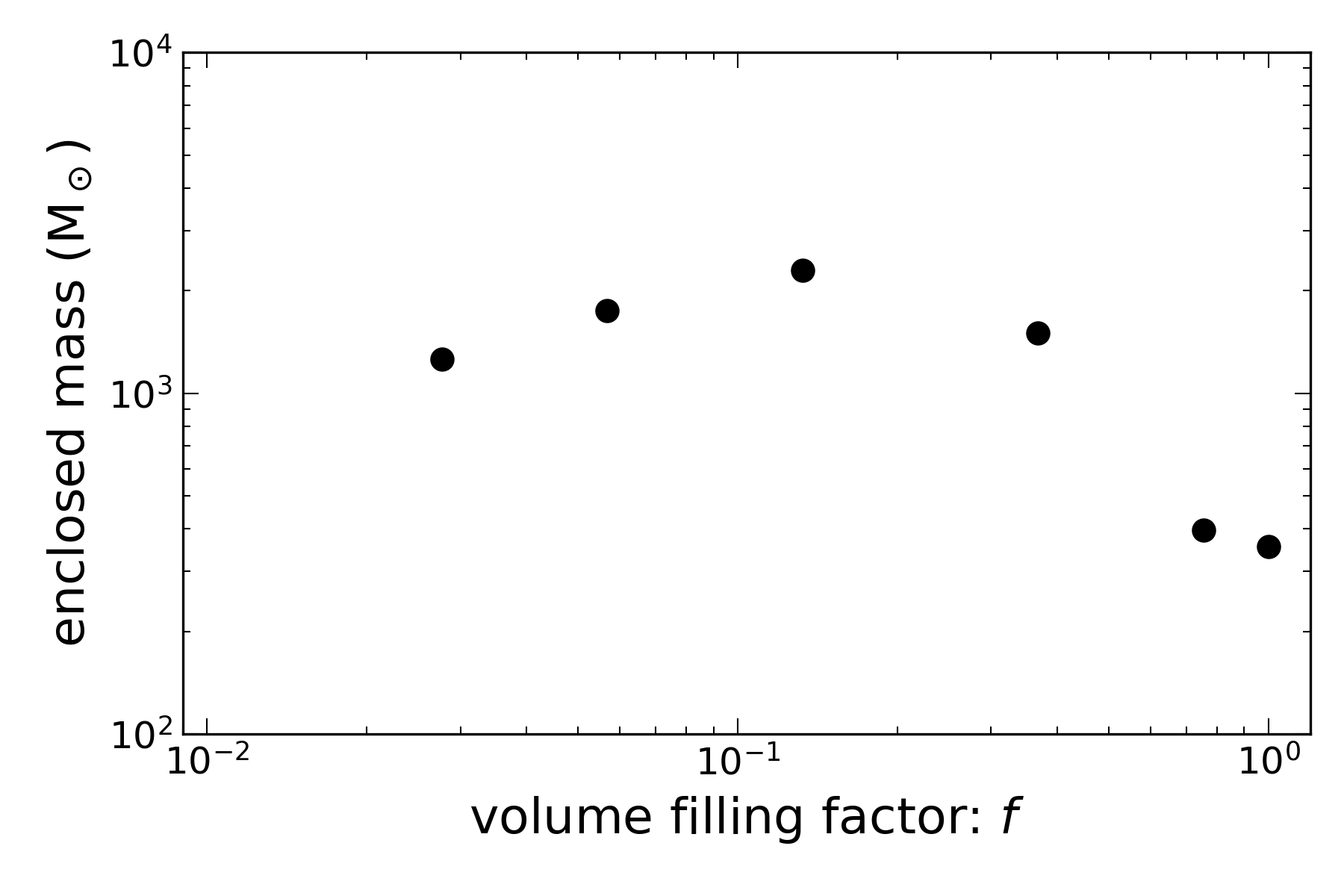}
  \caption{
  Dependence of the total gas mass enclosed within the forward shock on the volume filling factor. For a direct comparison with observations, the integrated values within the numerical domain are multiplied by a factor of eight to represent the equivalent values for the full sphere.}
  \label{fig:fV_vs_Mshell}
\end{figure}

To evaluate whether the required clumpy environment ($f \lesssim 0.1$) is physically compatible with observations of RX J1713.7$-$3946, we analyze the mass distribution of the surrounding interstellar gas.
In the case of RX J1713.7$-$3946, approximately $10^4\ M_{\odot}$ of total target gas mass exists in its vicinity \citep{Fukui2012ApJ74682}. In our simulations at $t = 1000\ \mathrm{yr}$, the total gas mass contained within the forward shock is plotted as a function of $f$ in Figure~\ref{fig:fV_vs_Mshell}. In all examined cases, this mass is significantly smaller than the observed value of $10^4\ M_{\odot}$. This discrepancy suggests that the majority of the observed gas has been swept up by either the \ion{H}{2} region expansion or stellar winds. Consequently, most mass resides within a wind/bubble shell located outside the forward shock, which does not emit gamma-rays, as discussed by \citet{InoueYamazakiInutsukaFukui2012ApJ74471}.

One might question whether the progenitor massive star of RX J1713.7$-$3946 could expel such a large amount of gas from its surroundings. However, it is well established that the expansion of H II regions and stellar wind bubbles driven by massive stars can easily reach scales of ~10 pc or more \citep[e.g.,][]{1975ApJ...200L.107C}. The mass of molecular cloud gas with a number density of 100 cm$^{-3}$ evacuated from a sphere with a 10-pc radius amounts to roughly $10^4\ M_{\odot}$. Furthermore, detailed radiation hydrodynamic simulations demonstrate that stars with masses of roughly $20\ M_{\odot}$ or more exert significant feedback effects on the surrounding gas of $10^4\ M_{\odot}$ or more via radiation \citep{2015A&A...580A..49I}. Therefore, the inferred structure around RX J1713.7$-$3946 is physically reasonable.

\begin{figure}[htbp]
  \centering
  \includegraphics[keepaspectratio, scale=0.5]{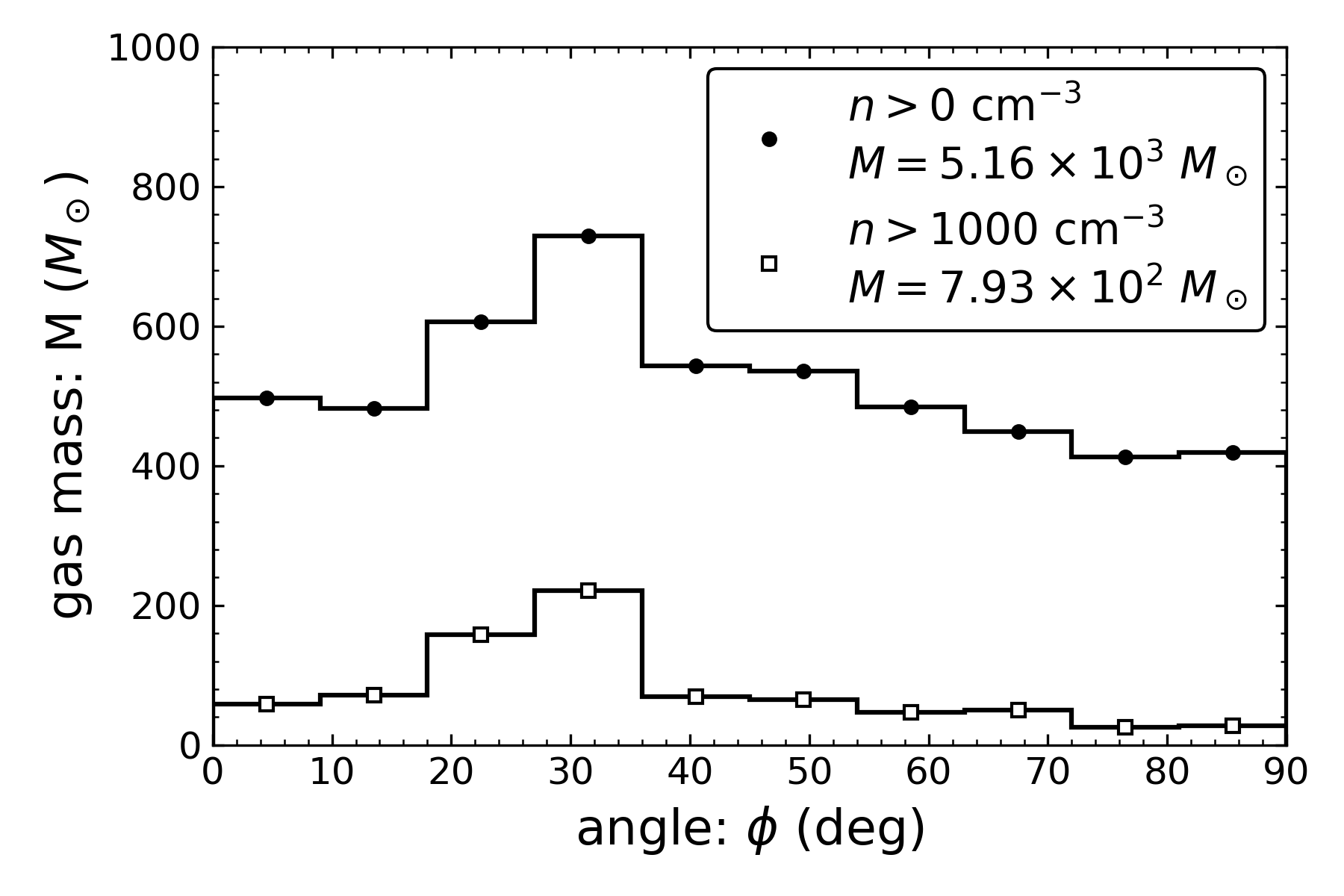}
  \caption{Azimuthal distribution of the initial molecular cloud mass (upper line) and that of the high-density components ($n > 10^3\ \mathrm{cm^{-3}}$) only (lower line). The total mass of the entire gas is $5.16 \times 10^3\ M_{\odot}$, and the total mass of the dense gas with $n > 10^3\ \mathrm{cm^{-3}}$ is $7.93 \times 10^2\ M_{\odot}$.}
  \label{fig:mass_vs_angle}
\end{figure}

In an environment where most of the gas is swept up into a wind/bubble shell, can we still observe the correlation reported by \citet{Fukui2012ApJ74682}? Specifically, is it possible to see the correlation between the azimuthal distribution of gamma-rays and the total gas mass around the SNR?
If the vast majority of the gas is located within the shell, one might argue that the mass of the clumps inside the forward shock—which are responsible for the hadronic gamma-ray emission—does not necessarily have to correlate with the total gas mass along that specific line of sight.
To address this issue, we present the azimuthal distribution of the gas mass for densities $n > 1000\ \mathrm{cm^{-3}}$ alongside the total gas mass across the entire density range in Figure~\ref{fig:mass_vs_angle}.
To calculate this azimuthal distribution, we analyze the column density data within the region of $r \equiv \sqrt{x^2+y^2} \le 10\ \mathrm{pc}$, obtained by integrating the turbulent molecular cloud density generated in Section~2.1 along the $z$-direction.

The figure clearly demonstrates a strong correlation between the high-density gas mass and the total gas mass. This reflects an intrinsic hierarchical property of the molecular cloud structure \citep[e.g.,][]{1981MNRAS.194..809L,1998A&A...336..697S}, where regions with a high abundance of dense clumps are naturally surrounded by a higher amount of diffuse gas as well. Therefore, even if most of the gas around RX J1713.7$-$3946 had been swept up into a wind shell beforehand, the azimuthal distributions of the gamma-ray emission and the total proton mass would still exhibit a tight correlation, consistent with the observational findings of \citet{Fukui2012ApJ74682}.

\section{Summary and Conclusions}
In this study, we have investigated the evolution of SNRs within a clumpy molecular cloud environment shaped by supersonic turbulence using three-dimensional hydrodynamic simulations.
We have modeled the pre-supernova environment by replacing the gas below a certain threshold number density with low-density hot gas.
In particular, we have characterized the ``clumpiness'' of the ambient medium by its volume filling factor, quantified its relationship with the propagation velocity of the forward shock, and discussed the implications for the SNR RX J1713.7$-$3946. Our primary conclusions are summarized as follows:

\begin{enumerate}
\item In a highly clumpy medium, the shock wave propagates preferentially through low-density regions while avoiding high-density clumps, allowing the remnant to expand while maintaining high spherical symmetry as a whole. Conversely, an increase in the number of high-density clumps subjects the shock wave to significant deceleration.

\item At $t = 1000\ \mathrm{yr}$, the forward shock velocity exhibits high sensitivity to the volume filling factor $f$. For cases with $f \lesssim 0.1$, the forward shock velocity evolves without significant deviation from the case where the entire upstream region is filled with hot gas. Therefore, to remain consistent with the fast forward shock velocity reported for RX J1713.7$-$3946, the clumpiness of the surrounding medium is constrained to $f \lesssim 0.1$.

\item Assuming hadronic gamma-ray emission for RX J1713.7$-$3946 and applying the constraint $f \lesssim 0.1$, the resulting decrease in the target gas mass relative to the $f = 1$ case (where all the gas acts as the target) suggests that the total energy of cosmic-ray protons is at least $W_{\rm tot} \gtrsim 10^{49}\ \mathrm{erg}$. This corresponds to more than 1\% of the typical supernova explosion energy ($10^{51}\ \mathrm{erg}$), which is consistent with the standard framework of cosmic-ray acceleration in SNRs.

\item However, for $f \lesssim 0.1$, the gas mass enclosed within the forward shock front becomes smaller than the total surrounding gas mass inferred from observations. This discrepancy indicates that the majority of the observed gas is swept up and resides outside the forward shock as a shell of an \ion{H}{2} region or a stellar wind bubble, implying that only a fraction of the total mass (the clump component inside the forward shock front) contributes to the gamma-ray emission.
\end{enumerate}

Our study supports the scenario that SNRs are the main acceleration sources of Galactic cosmic rays, although a significant question regarding their maximum achievable energy remains unanswered. This is because RX J1713.7$-$3946 is widely acknowledged to not be a PeVatron, at least at the present epoch \citep{2007A&A...464..235A,2008ApJ...685..988T,2018A&A...612A...6H,2024ApJ...965..113I}. 
While recent theoretical and observational studies suggest that SNRs at even younger stages are promising candidates for PeVatrons \citep{2013MNRAS.435.1174S,2021ApJ...922....7I,2021NatAs...5..460T,2026arXiv260422621C}, further investigation is required to determine whether they could have accelerated a sufficient amount of PeV cosmic rays during their earlier phases to remain consistent with the observed cosmic-ray spectrum.

\begin{acknowledgments}
We thank H. Susa, Y. Fukui, and H. Sano for fruitful discussions that significantly advanced this work.
Numerical computations were carried out on the Cray XD2000 system at the Center for Computational Astrophysics, National Astronomical Observatory of Japan.
The computations were also carried out on the Yukawa-21 supercomputer system at the Yukawa Institute for Theoretical Physics, Kyoto University.
This work was supported by Grants-in-Aid for Scientific Research from the Ministry of Education, Culture, Sports, Science and Technology (MEXT) of Japan under Grant Nos. 23H00129 and 26H02068 (TI).
KT was supported by JST SPRING, Grant Number JPMJSP2117.
\end{acknowledgments}

\bibliography{article}

\end{document}